# Electrical detection of current generated spin in topological insulator surface states: Role of interface resistance

C. H. Li,[1*] O. M. J. van 't Erve,[1] C. Yan,[2] L. Li,[2] and B. T. Jonker[1]

[1]Materials Science and Technology Division, Naval Research Laboratory, Washington, DC 20375, USA
[2]Department of Physics and Astronomy, West Virginia University, Morgantown, WV 26506 USA.

## Abstract

Current generated spin polarization in topological insulator (TI) surface states due to spin-momentum locking has been detected recently using ferromagnet/tunnel barrier contacts, where the projection of the TI spin onto the magnetization of the ferromagnet is measured as a voltage. However, opposing signs of the spin voltage have been reported, which had been tentatively attributed to the coexistence of trivial two-dimensional electron gas states on the TI surface which may exhibit opposite current-induced polarization than that of the TI Dirac surface states. Models based on electrochemical potential have been presented to determine the sign of the spin voltage expected for the TI surface states. However, these models neglect critical experimental parameters which also affect the sign measured. Here we present a Mott two-spin current resistor model which takes into account these parameters such as spin-dependent interface resistances, and show that such inclusion can lead to a crossing of the voltage potential profiles for the spin-up and spin-down electrons within the channel, which can lead to measured spin voltages of either sign. These findings offer a resolution of the ongoing controversy regarding opposite signs of spin signal reported in the literature, and highlight the importance of including realistic experimental parameters in the model.

* Corresponding author. Email: connie.li@nrl.navy.mil

## Introduction

Spin-momentum locking is one of the most remarkable properties of 3D topological insulators (TIs), where the spin and momentum of the carriers in the topologically protected surface states lie in-plane and are locked at right angles to each other[1-5]. This dictates that an unpolarized charge current induces a spontaneous spin polarization of known orientation (Fig. 1a). We recently demonstrated electrical detection of bias current generated spin polarization in TI surface states, where the projection of the TI spin onto the magnetization of a ferromagnet/tunnel barrier detector contact was detected as a voltage[6]. This potentiometric method has been adapted to measure the current generated spin in other TI systems[7-15]. However, conflicting signs of the measured spin voltage signals have been reported, as reflected in whether a high or low voltage signal is measured when the magnetization of the detector contact is parallel or antiparallel to the induced spin[6-15].

These discrepancies could be potentially attributed to the coexistence of a two-dimensional electron gas (2DEG) on the TI surface due to band bending, which may exhibit an opposite current induced polarization than that of the TI Dirac surface states[16]. Comparative measurements using the same ferromagnet/tunnel barrier detector contacts and identical measurement geometries carried out on InAs(001) reference samples where only 2DEG is expected indeed reveal opposite signs of the current induced spin for the InAs and $Bi_2Se_3$[14]. A potential complication to this control experiment is that the measured spin voltage arising from the trivial 2DEG states is also sensitive to the sign and value of the Rashba spin–orbit coupling parameter alpha[16], which can vary depending on the nature of the interface in a heterostructure[17]. However, positive values of alpha have been reported for various types of TI and the InAs(001) surface in the literature[18-22], suggesting that the discrepancies noted above arise from mechanisms of a different origin.

Models based on electrochemical potential have also been presented to derive the sign of the spin voltage that would be expected for the TI surface states[9,11,14]. However, these models only consider the spin-dependent electrochemical potential for the spin-up and spin-down electrons in the TI channel, and do not take into account key experimental parameters such as the interface resistance.

Here we present a Mott two-spin current resistor model[23] that takes into account such parameters. The model is based upon two parallel channels for spin-up and spin-down electrons,[24,25] and importantly includes contact and interface resistances at the current injecting contacts. We show that inclusion of interface resistances can cause a crossing of the voltage potential profiles of the spin-up and spin-down electrons along the channel, which can lead to measured spin voltages of either sign regardless of the spin polarization. These results demonstrate that the interpretation of electrical measurement of current-generated spin in TI surface states is more complex than previously considered, and that spin dependent resistances in both the channel and interfaces must be considered to correctly interpret the sign of the spin voltage measured.

## Results and Discussions

**Model.** The electrical detection of current-generated spin using a ferromagnetic detector is typically modeled as a simple 3-terminal geometry similar to that from Hong *et al.*[16] (Fig. 1b). Here the left contact is defined as the positive terminal, and the right contact as the negative or reference terminal. The +y direction is defined as the positive magnetic field direction (and ferromagnetic detector magnetization), and +x the direction of a positive (hole) current. For a positive hole current flowing through the TI surface states in the +x direction, the electrons flow from right to left in the -x direction, generating a spin orientation in the +y direction within the TI channel. In the models presented in Ref. 9 (Figs. 1d&e) and 11 (Figs. 3b&d), which we adopted in our own previous work (Ref. 14, Fig. 5), this splitting in the electrochemical potential (or spin voltage) for the spin-up and down electrons is simply represented by a pair of parallel linear profiles throughout the TI channel, which converge discontinuously (shown by a vertical line for one or both of the spin channels) at the current terminals. We find that this simple picture does not correctly represent the real experimental conditions, as critical parameters such as interface resistances are not taken into account.

Specifically, the interface resistances at the current injecting contacts are not necessarily symmetric due to their nonlinear nature. This is a result of a blanket layer of tunnel barrier material such as $Al_2O_3$ that is often deposited on the TI as the first step (for capping purposes and/or to simplify fabrication processes)[6,7,9-15]. It is therefore not only present at the ferromagnet/tunnel barrier spin detection contacts, but also at the interfaces of the current injecting contacts. Fig. 1c shows a typical *I-V* curve taken at 8 K between two $Au/Al_2O_3/Bi_2Se_3$ contacts of different sizes, showing the nonlinear nature of these contacts. This *I-V* characteristic indicates a non-negligible and nonlinear interface resistance where a voltage drop can be supported.

Interface resistances can be measured using transmission line measurements with a series of equally sized contacts separated by different distances, where the y-intercept of the plot of resistance versus contact separation is an indication of two times the interface resistance of the contacts. Alternatively, both four-probe and two-probe resistance measurements can be carried out, where the former is a measure of the channel resistance and the latter the sum of the channel resistance and two interface resistances.

In a typical experiment to electrically detect current generated spin in a TI thin film, an electron current flowing from right to left in the -x direction through the TI surface states establishes a net spin-up polarization (along +y direction) on the top surface of the TI film, and spin-down (along -y) polarization on the bottom surface. While these two physically separated spin channels are equivalent as required by time reversal symmetry, our spin detector contacts are nonetheless only on the top surface, and therefore only sensitive to the spins on the top surface. Hence the discussion and model herein pertain only to the spins on the top surface. Furthermore, calculations including spin-orbit coupling indicate that spin is no longer a good quantum number[26]. Thus while the current induced spin indeed has in-plane anisotropy, the polarization is momentum dependent with an average value reduced to only ~0.5 [26].

Shown in Fig. 1d is a schematic of our resistor circuit model for both spin-up and spin-down electrons traveling in two independent channels from the right to the left electrode on the top surface. Each component of the circuit, including the contacts and interfaces, is modeled as a resistor[23-25]. We have used a similar approach to model the spin filtering effects in graphene/ferromagnet magnetic tunnel junctions[27]. As electrons travel from the right gold electrode to the left, several resistances are encountered, (from right to left): resistance of the right Au electrode $R_{Au,R}$, resistance at the right Au/Al$_2$O$_3$/TI interface $R_{int,R}$, TI channel resistance $R_{TI}$, resistance at the left TI/Al$_2$O$_3$/Au interface $R_{int,L}$, and resistance of the left Au electrode $R_{Au,L}$. Some of these resistances will also be spin-dependent, as discussed below, and depending on their relative magnitudes, the voltage potential profile can vary significantly.

For electrons traveling from the right Au electrode in a steady state bias current, the resistance of the Au electrode is low for both spin-up (+y) and spin-down (-y) electrons. However, the interface resistance for spin-up and spin-down electrons entering into the TI channel may be different depending on their alignment with the states in the TI[28]. A left-flowing electron current in the TI surface states creates a spontaneous spin-up orientation (+y) due to spin-momentum locking. Hence for spin-up electrons entering into the top surface of the TI channel, this interface resistance will be lower since they align with the those in the TI surface states under this steady state condition. The opposite is true for spin-down electrons (-y) – the interface resistance will be higher due to their antiparallel alignment. Finally, as these electrons enter into the left Au electrode, the interface resistance will be similar for both spins since there are equal number of spin-up and spin-down states in the Au, i.e., the interface resistance here will not likely be spin-dependent. Similarly, the resistance of the left Au electrode for both spins will be the same and small.

Given that the overall voltage drop for both the spin-up and down channels must be the same across the left and right Au electrodes, and that the spin-up channel is clearly a lower resistance channel on the top surface, the current flowing through the spin-up channel ($I_\uparrow$) will be greater than that for the spin down channel ($I_\downarrow$), or $I_\uparrow > I_\downarrow$.

**Including interface resistances (assuming no spin-dependency).** In the simplest case, we take into account the interface resistances, but not their spin dependencies, i.e., the interface resistance is the same for both spin-up and down channels, or $R_{int,R\uparrow} = R_{int,R\downarrow}$ (for the right $Au/Al_2O_3/TI$ interface). Here, due to the greater current in the spin-up channel (blue), $I_\uparrow > I_\downarrow$, the voltage drop at the interfaces is greater for the spin-up ($V_{int,R\uparrow}$) than the spin-down ($V_{int,R\downarrow}$) channel, as depicted by the steeper slope for the blue lines (spin-up) within the right $Au/Al_2O_3/TI$ interface region in Fig. 2a, and a smaller slope for the red lines (spin-down). The same situation is depicted for the left $Au/Al_2O_3/TI$ interface as well, as it is not a spin-dependent interface in any case for electrons entering into the Au electrode. Connecting the end points of the voltage profiles for spin-up (blue) and spin-down (red) channels yields the profile shown in Fig. 2a, where a crossing of the spin-up and down voltage profiles within the TI channel is evident. This crossing necessarily occurs due to the larger voltage drop for the spin-up channel (owing to higher current) at both interfaces, while the total voltage drop for both spin channels must remain the same. It is important to recognize that this crossing does not imply a change in sign of the spin polarization in the channel or direction of the charge flow, but will result in a reversal of the measured spin voltage loop, as discussed below. A positive or negative slope of the voltage potential indicates the direction of the flow of electrons.[24,25] Here the sign of the slope is constant and does not change, indicating that the current direction does not change. The magnitude of the slope is determined by the number of electrons and the resistance where it changes in different regions (e.g., channel, interface), indicating a change of resistance in those regions, but not a change in current direction or spin polarization.

Also note that a crossing would appear to be unexpected, where one would naturally expect a constant splitting between the spin-up and spin-down bands ($\Delta\mu_s$) for current generated spin. However, this would be true as an *equilibrium* condition for an *isolated* TI. As interface and interface resistance are a necessary components of any electrical transport measurement, they must be taken into account, which indeed modifies the potential profiles in the TI channel.

Specifically, charge carrier conversion at the interface creates boundary conditions that ensure the spin and current continuity across the interface. This results in the splitting of the spin up and spin down levels near the interface,[24,25] for example for a ferromagnet /normal metal junction. The equilibrium condition within the ferromagnet ($\Delta\mu_s>0$) and normal metal ($\Delta\mu_s=0$) are only observed beyond the spin diffusion length away from the interface (typically on the order of nms-µms). In the TI case, however, since the spin coherence length in the TI is very large (e.g., > 100's µms), the interface perturbation extends far into the channel, and the *equilibrium* conditions that would be expected for an isolated TI, i.e., a constant spin splitting between the spin-up and spin-down bands across the TI channel, are never realized.

The relative magnitudes of the interface and TI channel resistances would change the magnitude of the splitting between the spin-up and spin-down channels, as shown in Fig. 2b for a smaller interface resistance, where the overall voltage drops at the interfaces are smaller but the existence of a crossing is present nonetheless. This crossing indicates that the relative levels of the spin-up and spin-down levels in the voltage profile are not uniform across the TI channel, but in fact reverse, and may lead to either sign of the spin voltage measured, as discussed in further detail below.

**With spin-dependent interface resistances.** Next we consider an interface resistance that is spin-dependent. Again with a left-flowing electron current through the TI surface states, spin-up states are generated due to spin-momentum locking. Hence, at the right $Au/Al_2O_3/TI$ interface, the spin-up electrons entering into the TI channel will encounter a lower interface resistance than that of spin-down electrons, i.e., $R_{int,R\uparrow}<R_{int,R\downarrow}$. And since the current through the spin-up channel is greater, $I_\uparrow>I_\downarrow$, the voltage drop at the interface for spin-up and spin-down electrons ($V_{int,R\uparrow}$ and $V_{int,R\downarrow}$, respectively) can have two different outcomes: $V_{int,R\uparrow}<V_{int,R\downarrow}$ or $V_{int,R\uparrow}>V_{int,R\downarrow}$ (Fig. 3a and b, respectively), depending on the relative magnitudes of the currents through the spin-up and down channels ($I_\uparrow$, $I_\downarrow$), compared to that of the spin-dependent resistances at the interface ($R_{int,R\uparrow}$ and $R_{int,R\downarrow}$). In the case that $V_{int,R\uparrow}<V_{int,R\downarrow}$, (Fig. 3a, due for example to $I_\uparrow\geq I_\downarrow$, $R_{int,R\uparrow}<<R_{int,R\downarrow}$), no crossing occurs along the channel, with the spin-down band in Fig. 3a remaining above the spin-up band. However, in the case that $V_{int,R\uparrow}>V_{int,R\downarrow}$, (Fig. 3b, due to for example $I_\uparrow>>I_\downarrow$, $R_{int,\uparrow}\leq R_{int,\downarrow}$), a crossing is clearly produced. Note that the left $TI/Al_2O_3/Au$ interface is still spin-*in*dependent for both spin-up and down electrons entering into the left Au electrode, or $R_{int,L\uparrow}=R_{int,L\downarrow}$, and since $I_\uparrow>I_\downarrow$, the voltage drop for spin-up is still greater than that of the spin down at the left $TI/Al_2O_3/Au$ interface.

Clearly the current injecting interface is an integral component of these circuit diagrams and the voltage drop at these interfaces must be considered. The inclusion of these interface resistances can create a crossing of the voltage profiles of the spin-up and spin-down electrons, and can lead to either sign of the measured spin voltage depending on the details of the spin-dependent resistances at the interface and channel.

**Expected line shape measured at different detector contacts.** A ferromagnetic detector contact is used to probe the spin-up and spin-down voltage profiles. An external magnetic field is applied to orient the magnetization of the ferromagnet. However, the magnetic moment of a ferromagnetic metal is opposite to the orientation of its majority spin[29]. Hence if the ferromagnetic detector exhibit *+M* magnetization (oriented along *+y*), its majority spin is oriented along *–y*, and will probe the spin-down electrons ($V_\downarrow$) in the TI channel. Conversely, *-M* magnetized detector probes spin-up levels ($V_\uparrow$).

For the voltage profiles shown in Fig. 3a (and reference being the right electrode), the spin-down voltage level probed by *+M* magnetization is *V(+M)=($V_\downarrow$-$V_R$)*, and the spin-up voltage level probed by *-M* magnetization is *V(-M)=($V_\uparrow$-$V_R$)*. Since the spin-down level (red) is above spin-up (blue), or $V_\downarrow$>$V_\uparrow$, then *V(+M)>V(-M)*, this produces a high voltage signal for positive magnetic field (the detector magnetization is parallel to the TI spin (spin-up)), and a low voltage at negative field, when the magnetization is antiparallel to the TI spin, as illustrated by the hysteresis loop shown in Fig. 3c. This sign is consistent with the observations in Refs. 9,11,13. Note that the relative high and low signals are not affected by a simple linear background subtraction and centering around the vertical axis.

Similarly for the voltage profiles shown in Fig. 3b in the center of the channel where the spin-up level (blue) is above the spin-down (red), or $V_\uparrow$>$V_\downarrow$, then *V(-M)>V(+M)*, yielding a low voltage signal at the positive field, and a high voltage at negative field, as shown by the hysteresis loop in Fig. 3d. This is clearly inverted relative to that of Fig. 3c, and the measured voltage will be opposite in sign. This sign is consistent with the observations reported in Refs. 6,13-15. Note that to the left of the level crossing very near the left TI/Al$_2$O$_3$/Au interface, the spin-up level (blue) is below the spin-down (red), and the opposite sign of the spin signal, $\Delta V=V(+M)-V(-M)$, will be detected. Clearly, the measured sign of the spin voltage is directly dependent on the spin-dependent resistances at the interface and channel.

**Reversing the current direction.** Reversing the current direction, or electron motion, in the +x direction (from left to right electrode) gives rise to a spin-down orientation due to spin-momentum locking. Hence spin-down channel is the lower resistance channel, and $I_\downarrow>I_\uparrow$. For electrons entering from the left Au electrode into the TI channel, the interface resistance at the left Au/Al$_2$O$_3$/TI electrode is spin-dependent, while the right TI/Al$_2$O$_3$/Au interface is not. At the left interface, the spin-down electrons entering the TI channel encounter a lower interface resistance than the spin-up electrons, i.e., $R_{int,L\downarrow}<R_{int,L\uparrow}$. Again, since now $I_\downarrow>I_\uparrow$, the voltage drop at this interface for spin-up and spin-down electrons can have two different outcomes: $V_{int,L\downarrow}<V_{int,L\uparrow}$ (due for example to $I_\downarrow \geq I_\uparrow$, and $R_{int,L\downarrow}<<R_{int,L\uparrow}$) as shown in Fig. 4a, or $V_{int,L\downarrow}>V_{int,L\uparrow}$ (for $I_\downarrow>>I_\uparrow$, and $R_{int,L\downarrow}\leq R_{int,L\uparrow}$) as shown in Fig. 4b, where a crossing occurs resulting in the opposite alignment of the spin-up and spin-down voltage profile than that in Fig. 3b.

The expected magnetic field dependence of the voltages measured by a ferromagnetic detector is shown in Figs. 4c and d. These hysteresis curves are inverted relative to those of Figs. 3c and d, respectively, due to the reversed current direction, consistent with that expected from current induced spin polarization, and experimental observations[6-15].

**Rectifying interface resistance shifts the crossing to one side.** As noted above, the interface resistances at the left TI/Al$_2$O$_3$/Au and right Au/Al$_2$O$_3$/TI interfaces are not symmetric, because the interface resistance is spin dependent when entering the TI channel, and spin-*in*dependent when entering the Au electrode. Even though the TI is a semiconductor that supports metallic surface states, the metal/TI current injecting contacts are typically non-ohmic and/or rectifying, due to TI surface oxidation (metal contact deposition typically performed *ex situ*), and/or the inclusion of a tunnel barrier such as $Al_2O_3$ at the interface[6,7,9-15]. This is evident from the *I-V* curve in Fig. 1b showing rectifying behavior. This results in a junction where the magnitudes of these two interface resistances can vary depending on the current direction, *i.e.,* higher resistance entering into the TI channel, and lower resistance entering into the Au electrode. This is depicted by the larger voltage drop at the higher resistance interface (entering the TI channel) in Figs. 3a,b and 4a,b. This asymmetry leads to a larger splitting between the spin-up and spin-down voltage levels at the higher resistance interface, and therefore pushes the crossing towards the opposing end of the TI channel (Figs. 3b&4b). Hence, the spin signal probed at points along the TI channel may indeed be of the same sign, although a narrow detector contact placed very close to the opposite end of the TI channel (entirely on the opposing side of the crossing) would detect an opposite sign.

In summary, we have developed a more realistic model to derive the sign of the current-induced spin voltages on the top surface of a TI measured by a ferromagnet detector contact, that takes into account crucial experimental parameters such as interface resistances. In this Mott two-spin current resistor model, two parallel channels for spin-up and spin-down electrons are modelled separately, and we find that spin-dependent interface resistance at the current injecting contact plays an important role. Depending on the relative magnitudes of the currents through the spin-up and spin-down channels compared to that of the spin-dependent interface resistances, a crossing of the voltage profiles of the spin-up and spin-down electrons may occur, which can lead to measured spin voltages of either sign. These results reconcile conflicting reports in the literature, and further highlight the intricate nature of the seemingly straightforward electrical measurement of current generated spin in TI surface states, where real experimental parameters such as spin dependent resistances in both the channel and at current injecting interfaces must be considered to accurately account for the sign of spin voltage measured.

**Acknowledgement:** The authors acknowledge support from Naval Research Laboratory (NRL) core programs, and from the Department of Energy (DE-SC0017632) at the West Virginia University.

**Author contributions**

O.M.J.E. and C.H.L. developed the model. C.Y. and L.L. grew the TI films and performed surface characterizations. C.H.L. grew the $Fe/Al_2O_3$ tunnel barrier contacts, carried out device fabrication and transport measurements with assistance from O.M.J.E. C.H.L., O.M.J.E., and B.T.J. analyzed the results and wrote the paper. All authors discussed and commented on the manuscript.

**Competing interests**

The authors declare no competing financial and/or non-financial interests.

**Data availability.** The datasets generated during and/or analyzed during the current study are available from the corresponding author on reasonable request.

**FIGURE CAPTION**

**Fig. 1 Schematic of TI surface bands and model for experimental concept.** (a) Left panel: Dirac cone of the TI surface states (blue), with the spin at right angles to the momentum at each point. Right panel: Top view of a slice in the $k_x$-$k_y$ plane of the TI surface states. An applied current produces a net momentum along $k_x$ and spin-momentum locking gives rise to a net spin polarization oriented in-plane and at right angles to the current. (b) Schematic of a simple 3-terminal geometry for the potentiometric measurement of current generated spin in topological insulators. (c) Typical I-V curve taken at 8 K between a current injecting $Au/Al_2O_3/Bi_2Se_3$ contact and another $Au/Al_2O_3/Bi_2Se_3$ contact of different size, showing a nonlinear behavior. (d) Schematic of a resistor circuit model for spin-up and spin-down electrons traveling from the right to the left electrode, where each component of the circuit from the contacts to interfaces are modeled as a resistor.

**Fig. 2 Voltage profiles with non spin-dependent interface resistance.** Voltage profiles for the spin-up (blue) and spin-down (red) electrons at the $Au/Al_2O_3/TI$ current injecting contacts and within the TI channel for a left flowing current, assuming interface resistance is not spin-dependent, for (a) high interface resistance, and (b) low interface resistance.

**Fig. 3 Voltage profiles with spin-dependent interface resistance.** Voltage profiles for the spin-up (blue) and spin-down (red) electrons at the $Au/Al_2O_3/TI$ current injecting contacts and within the TI channel for a left flowing current, assuming interface resistance is spin-dependent, for the case (a) $V_{int,R\uparrow}<V_{int,R\downarrow}$ (due to for example $I_\uparrow \geq I_\downarrow$, $R_{int,R\uparrow}<<R_{int,R\downarrow}$), and (b) $V_{int,R\uparrow}>V_{int,R\downarrow}$ (due to for example $I_\uparrow>>I_\downarrow$, $R_{int,R\uparrow}\leq R_{int,R\downarrow}$). Predicted lineshape for the spin voltage measured by a ferromagnet/tunnel barrier detector contact for the voltage profiles in (a) and (b) are shown in (c) and (d), respectively.

**Fig. 4 Voltage profiles with spin-dependent interface resistance when reversing the current flow.** Voltage profiles for the spin-up (blue) and spin-down (red) electrons at the $Au/Al_2O_3/TI$ current injecting contacts and within the TI channel for a *right* flowing current, assuming interface resistance is spin-dependent, for the case (a) $V_{int,L\uparrow}<V_{int,L\downarrow}$ (due to for example $I_\uparrow \geq I_\downarrow$, $R_{int,L\uparrow} << R_{int,L\downarrow}$), and (b) $V_{int,L\uparrow} > V_{int,L\downarrow}$ (due to for example $I_\uparrow >> I_\downarrow$, $R_{int,L\uparrow} \leq R_{int,L\downarrow}$). Predicted lineshape for the spin voltage measured by a ferromagnet/tunnel barrier detector contact for the voltage profiles in (a) and (b) are shown in (c) and (d), respectively.

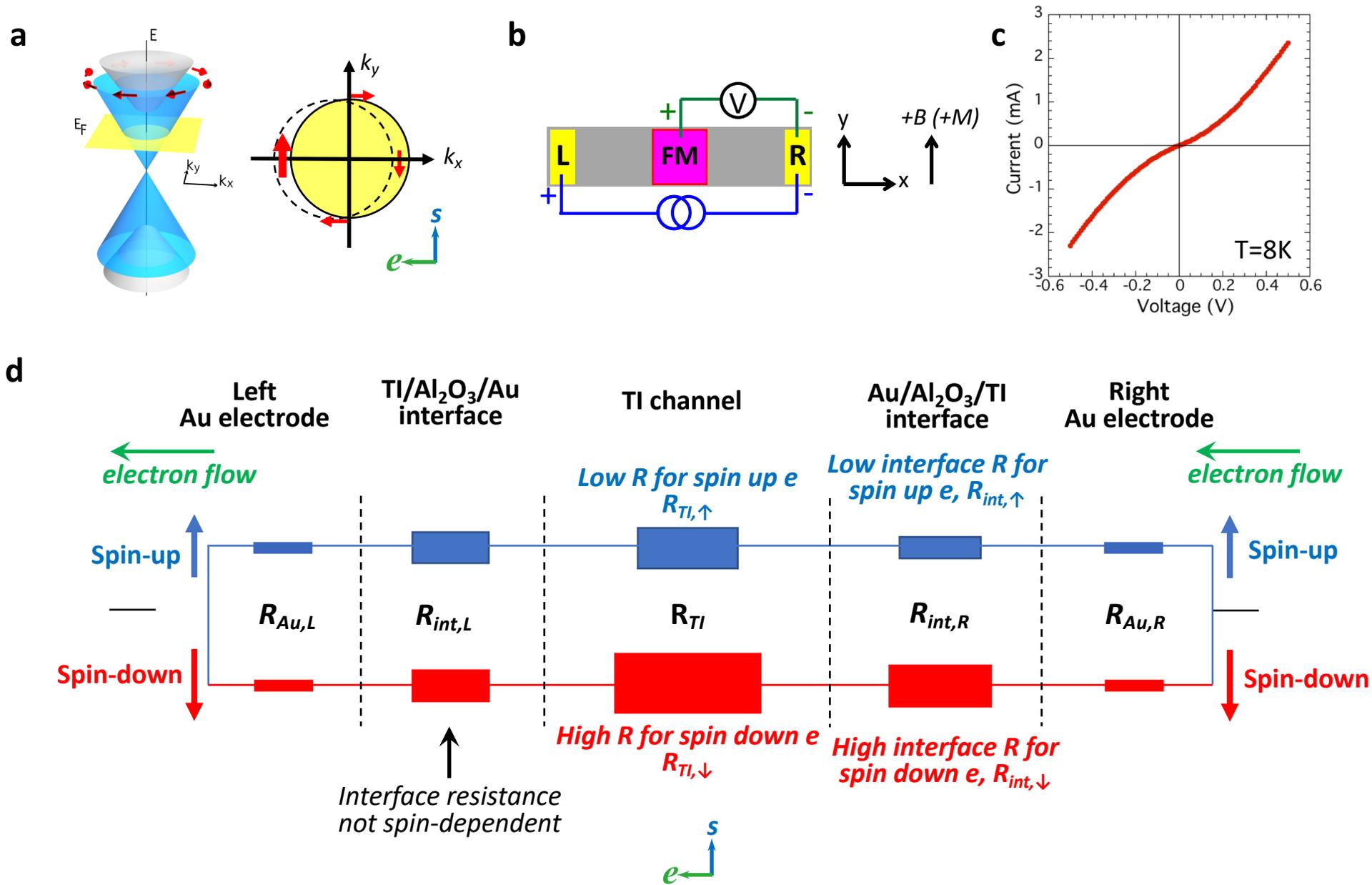

Li et al., Fig. 1

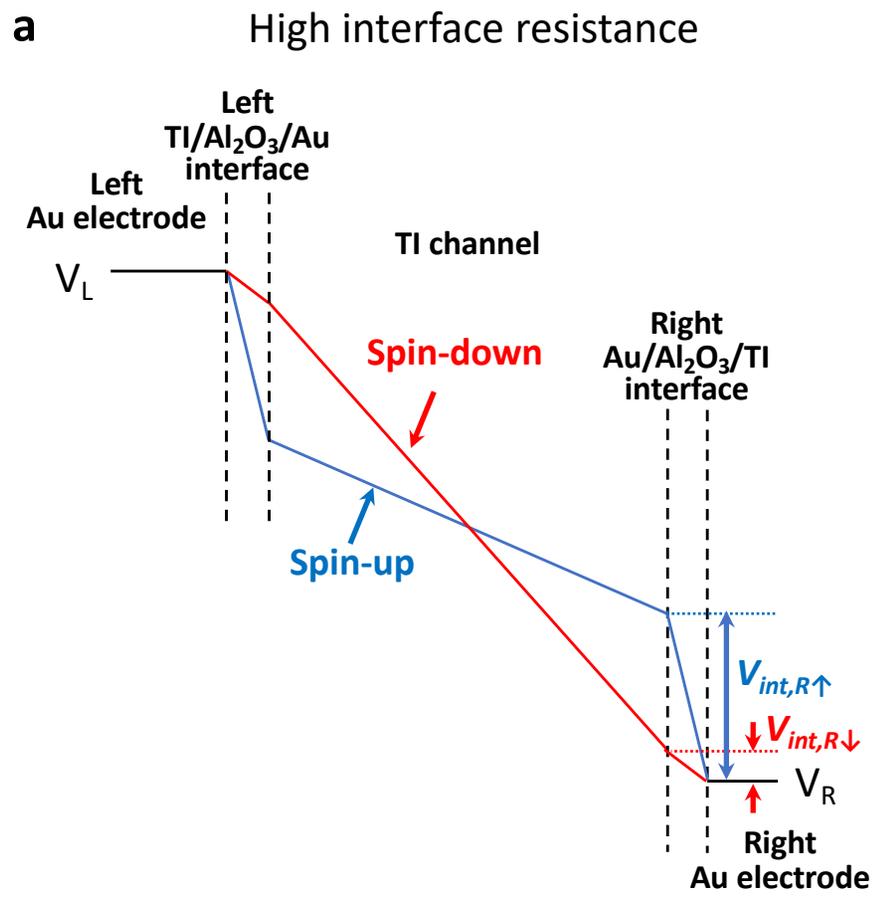 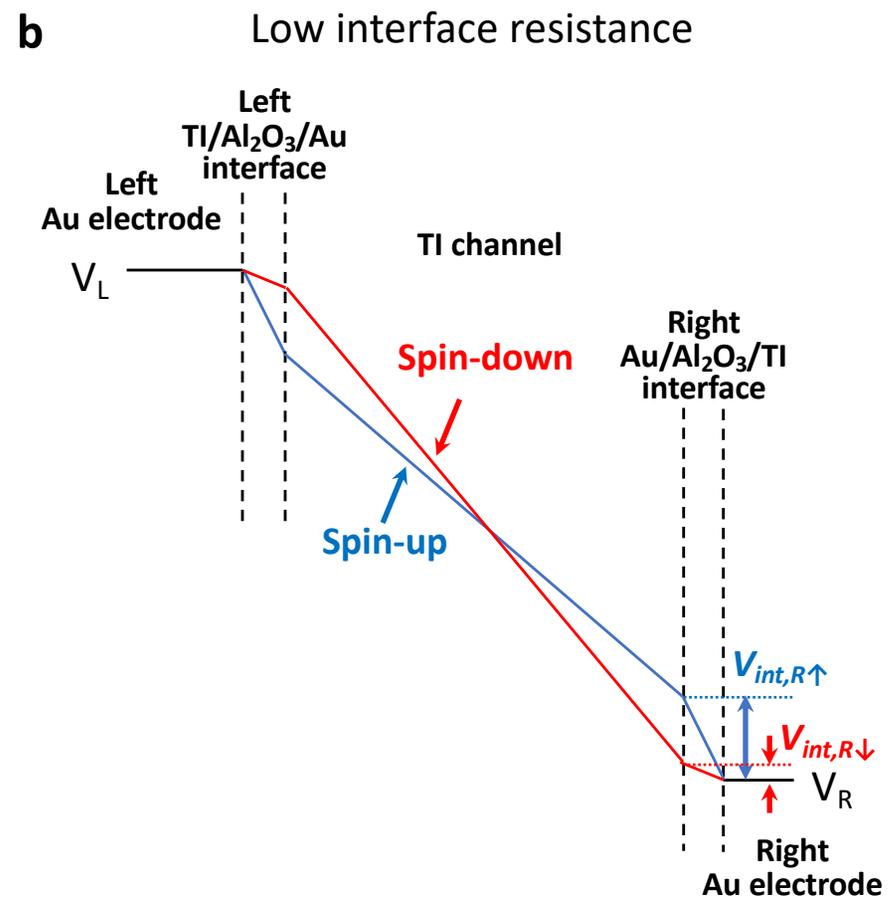

Li et al., Fig. 2

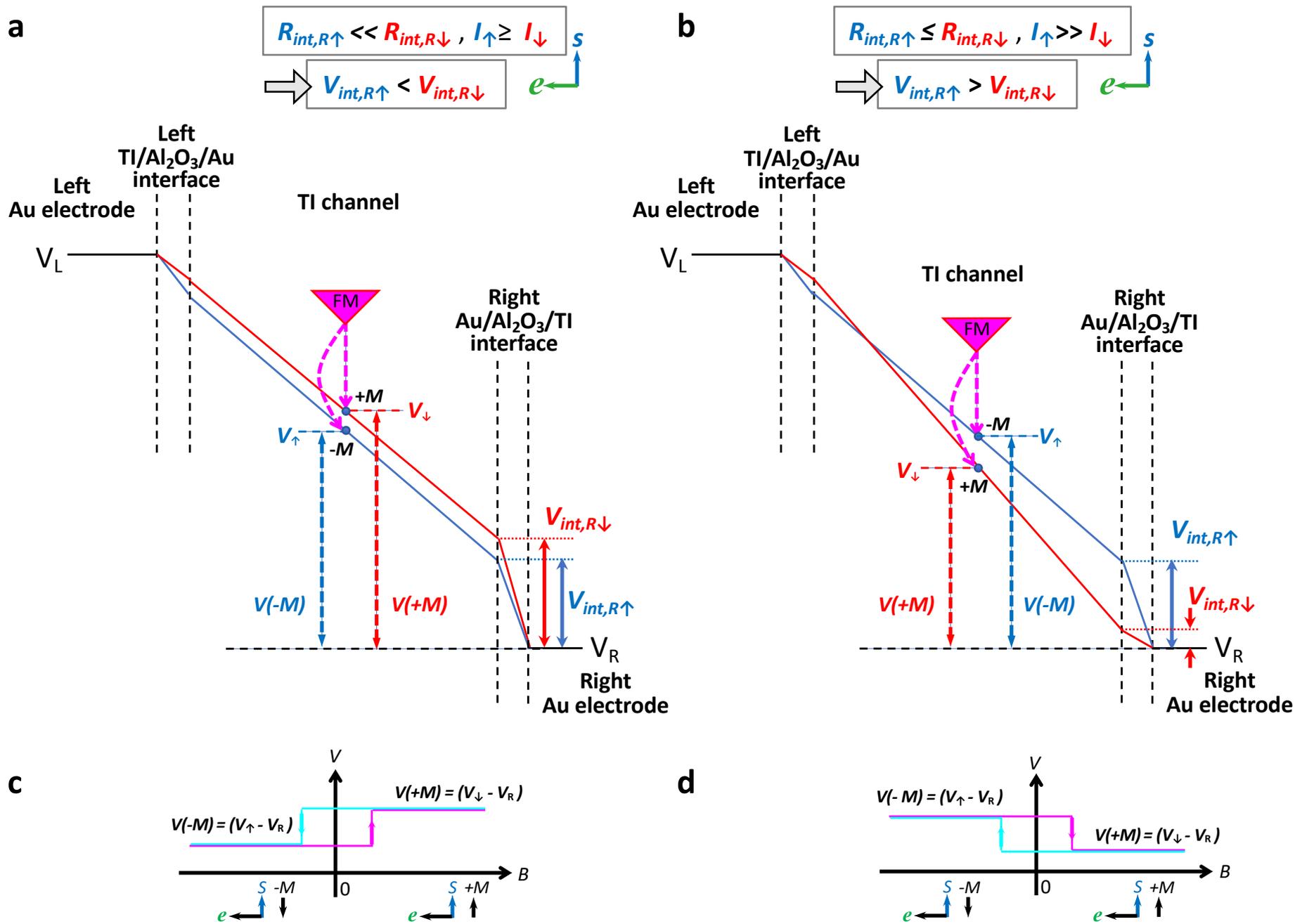

Li et al., Fig. 3

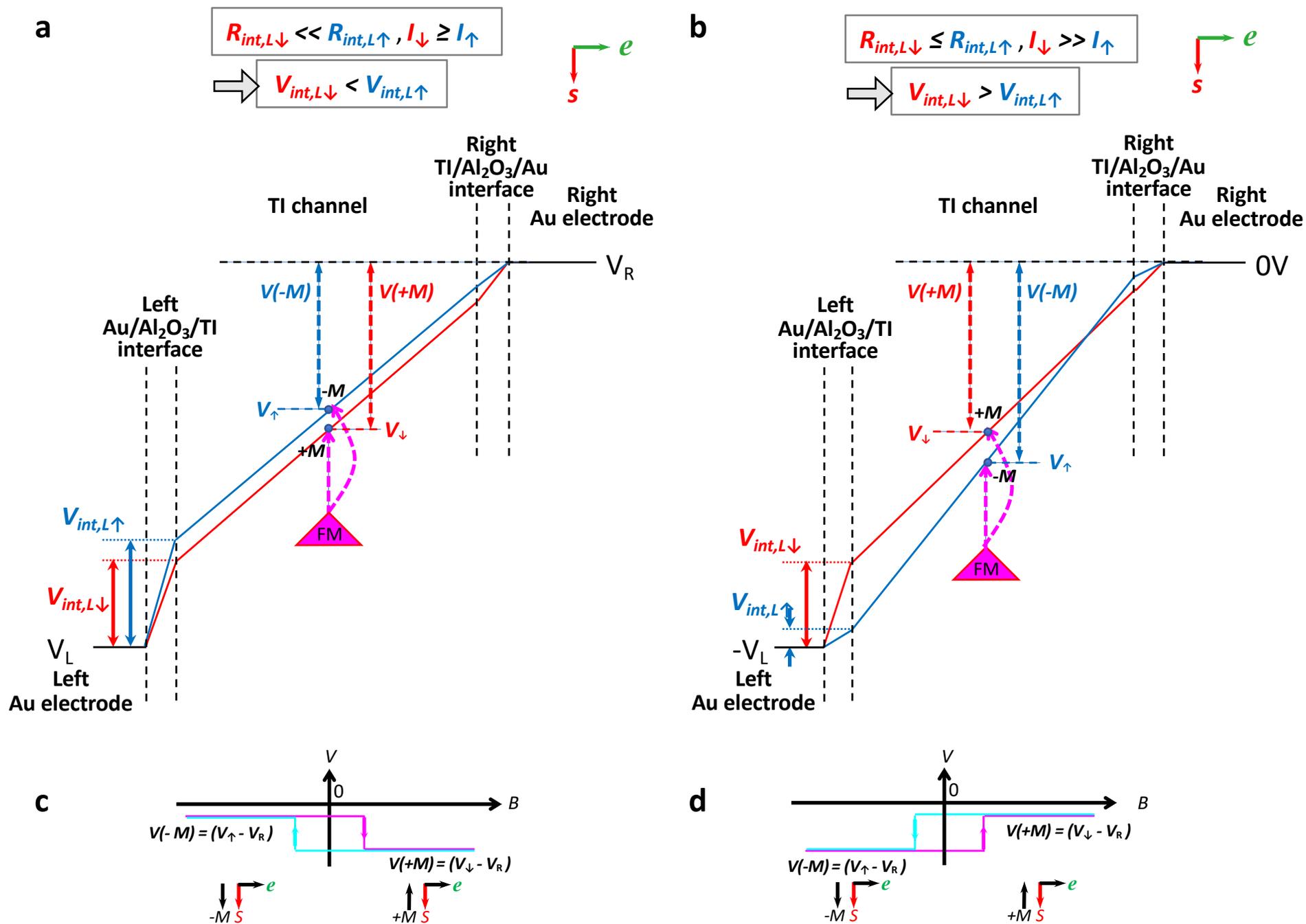

Li et al., Fig. 4